\font\teneufm=eufm10 
\documentclass[12pt]{article}

\usepackage{amsfonts}
\begin{document}

\def\Eh{\mbox{\teneufm\char 83}}

\title{Supersymmetry and discrete transformations of the Dirac 
operators in Taub-NUT geometry}

\author{Ion I. Cot\u aescu \thanks{E-mail:~~~cota@physics.uvt.ro}\\ 
{\small \it West University of Timi\c soara,}\\
       {\small \it V. P\^ arvan Ave. 4, RO-1900 Timi\c soara, Romania}
\and
Mihai Visinescu \thanks{E-mail:~~~mvisin@theor1.theory.nipne.ro}\\
{\small \it Department of Theoretical Physics,}\\
{\small \it National Institute for Physics and Nuclear Engineering,}\\
{\small \it P.O.Box M.G.-6, Magurele, Bucharest, Romania}}
\date{\today}

\maketitle

\begin{abstract}

It is shown that the $N=4$  superalgebra of the Dirac theory in Taub-NUT 
space has different unitary representations related among themselves through 
unitary $U(2)$ transformations. In particular the $SU(2)$ transformations are 
generated by the spin-like operators constructed with the help of the same 
covariantly constant Killing-Yano tensors which generate Dirac-type 
operators. A parity operator is defined and some explicit transformations 
which connect the Dirac-type operators among themselves are given.
These transformations form a discrete group which is a realization of 
the quaternion discrete group. The fifth  Dirac operator constructed 
using the non-covariant Killing-Yano tensor of the Taub-NUT space is 
quite special. This non-standard Dirac operator is connected with the 
hidden symmetry and is not equivalent to the Dirac-type operators of 
the standard $N=4$ supersymmetry.

Pacs 04.62.+v

\end{abstract}

\section{Introduction}

The theory of the usual or hidden symmetries of 
the Lagrangian quantum field theory on curved spacetimes, give rise to 
interesting mathematical problems concerning the properties of the physical  
observables. It is known that one of the largest algebras of conserved 
operators is produced by the Euclidean Taub-NUT geometry since beside usual 
isometries this has a hidden symmetry of the Kepler type \cite{GM,GRFH}. 
When discussing the geodesic equations in the Taub-NUT metric the 
existence of extra conserved quantities was noticed. These reflect a 
symmetry of the phase space of the system and enable the Schr\"odinger 
\cite{GM,CV1} and Dirac equations \cite{DIRAC,CH,CV2} to be separated 
in a special coordinate system. This is related to the existence of a 
St\"ackel-Killing tensor of rank $2$ in Taub-NUT space.
 
The theory of the Dirac equation in the Kaluza-Klein monopole field was 
studied in the mid eighties \cite{DIRAC}. An attempt to take into account 
the Runge-Lenz vector of this problem was done in \cite{CH}. We have 
continued this study showing that the Dirac equation is analytically 
solvable \cite{CV2} and determining the energy eigenspinors of the central 
modes. Moreover, we derived all the conserved observables of this theory, 
including those associated with the hidden symmetries of the Taub-NUT 
geometry. Thus we obtained the Runge-Lenz vector-operator of the Dirac 
theory, pointing out its specific properties \cite{CV3}. The consequences 
of the existence of this  operator were studied in \cite{CV4} showing that 
the dynamical algebras of the Dirac theory corresponding to different 
spectral domains are the same as in the scalar case \cite{GRFH} but 
involving  other irreducible representations. 

The Taub-NUT space is also of mathematical interest, the main features 
of the Taub-NUT metric relevant here are the fact that it is a $4$ 
dimensional hyper-K\"ahler  metric and possesses special tensors - 
St\"ackel-Killing and Killing-Yano tensors \cite{GRFH,H}. A hyper-K\"ahler
manifold is a Riemannian manifold modeled on a quaternion 
inner-product space. In fact a hyper-K\"ahler manifold is a manifold 
whose Riemannian metric is K\"ahler with respect to three different 
complex structures. In the $4$ dimensional case the holonomy group 
$Sp(1)\subset SO(4)$ is the same as $SU(2)\subset SO(4)$.

In the Taub-NUT geometry four Killing-Yano tensors are known to exist. 
Three of these are special because they are covariantly constant and 
define the complex structures of the manifold. Using these covariantly 
constant Killing-Yano tensors it is possible to construct new 
Dirac-type operators \cite{CL} which anticommute with the standard Dirac 
operator. The aim of this paper is to prove explicitly that these 
operators and the standard Dirac one are equivalent among themselves. 

We show that the representation of the whole theory can be changed using 
the $U(2)$ transformations among them the $SU(2)$ ones are generated just 
by the spin-like operators constructed using the above mentioned three 
Killing-Yano tensors \cite{CV2}. 
Based on these results, we define the parity transformation and a discrete 
group with eight elements formed by the transformations which relate to 
each other the four Dirac operators and their parity transformed as well. 
We show that this discrete group is a realization of the quaternion group 
which is isomorphic with the dicyclic group of order eight.  

The Taub-NUT space also possesses  a Killing-Yano tensor which 
is not covariantly constant. The corresponding non-standard operator, 
constructed with the general rule \cite{CL} anticommutes with the 
standard Dirac operator but is not equivalent to it. This non-standard 
Dirac operator is connected with the hidden symmetries of the space 
allowing the construction of a conserved vector operator analogous to 
the Runge-Lenz vector of the Kepler problem \cite{CV3}. The final objective 
here is to discuss the behavior of this operator under discrete 
transformations pointing out that the hidden symmetries are in some sense 
decoupled from the discrete symmetries studied here. 
The explanation of this distinction is that the standard $N=4$ 
supersymmetry  are linked to the hyper-K\"ahler structure of the 
Taub-NUT space. The corresponding supercharges close on the Hamiltonian 
of the theory. The quantal anticommutator of the Dirac-type operators 
closes on the square of the Hamiltonian operator. On the other hand, 
the non-standard supercharge involving the non-covariant Killing-Yano 
tensor does not close on the Hamiltonian. The appearance of the 
non-covariant Killing-Yano tensor in this context is not surprising 
since it also plays an essential role in the existence of hidden 
symmetries. Its existence requires the Weyl tensor to be of Petrov type 
$D$. The quantal anticommutator of the non-standard Dirac operator does 
not close on the square of the Hamiltonian, as would Dirac-type 
operators, rather on a combination of different conserved operators of 
the theory.

The paper is organized as follows.
In the next two sections we introduce the first four Dirac operator which 
constitute the $N=4$ superalgebra and we define the transformations leading 
to equivalent representations of the whole theory. These allow us to extract 
in section 4 the discrete transformations which show that these Dirac 
operators are equivalent among themselves. The role of the fifth Dirac 
operator is briefly discussed in section 5. Section 6 contains some 
discussion.

%We use natural units with $\hbar=c=1$.

\section{Dirac operators of the Taub-NUT space}

Let us consider the Taub-NUT space and the chart with Cartesian coordinates 
$x^{\mu}$  ($\mu, \nu,...=1,2,3,4$) having the line element  
\begin{equation}\label{(met)} 
ds^{2}=g_{\mu\nu}dx^{\mu}dx^{\nu}=\frac{1}{V}dl^{2}+V(dx^{4}+
A_{i}dx^{i})^{2}\,,
\end{equation}   
where $dl^{2}=(d\vec{x})^{2}=(dx^{1})^{2}+(dx^{2})^{2}+(dx^{3})^{2}$
is the  Euclidean three-di\-men\-sio\-nal line element and $\vec{A}$ is 
the gauge field of a monopole. Another chart suitable for applications is 
that of spherical coordinates, $(r,\,\theta,\,\phi,\,\chi)$, among them 
the first three are the  spherical coordinates commonly associated with 
the  Cartesian space ones, $x^{i}$  ($i,j,...=1,2,3$), while 
$\chi+\phi=- x^{4}/\mu$. The real number $\mu$ is the  parameter of the 
theory which enters in the form of the function $1/V(r)=1+\mu/r$. 
The unique non-vanishing component of the vector potential in 
spherical coordinates is $A_{\phi}=\mu(1-\cos\theta)$. 
This space has the isometry group  $G_{s}=SO(3)\otimes U(1)_4$ formed by 
the rotations of the Cartesian space coordinates and $x^{4}$ translations. 
The $U(1)_4$ symmetry is important since this eliminates the so called NUT 
singularity if $x^4$ has the period $4\pi\mu$. 

For the theory of the Dirac operators in Cartesian charts of the Taub-NUT 
space, it is convenient to consider the local frames given by tetrad fields 
$e(x)$ and $\hat e(x)$ as defined in \cite{P} while the four Dirac matrices 
$\gamma^{\hat\alpha}$, that  satisfy 
$\{ \gamma^{\hat\alpha},\, \gamma^{\hat\beta} \} 
=2\delta^{\hat\alpha \hat\beta}$, have to be written in the following  
representation 
\begin{equation}\label{(gammai)} 
\gamma^i = -i
\left(
\begin{array}{cc}
0&\sigma_i\\
-\sigma_i&0
\end{array}\right)\,,\quad  
\gamma^4 =
\left(
\begin{array}{cc}
0&{\bf 1}_2\\
{\bf 1}_2&0
\end{array}\right)\,,
\end{equation}
where all of them are self-adjoint. In addition we consider the matrix
\begin{equation}\label{(gamma5)} 
\gamma^5 = \gamma^1\gamma^2\gamma^3\gamma^4 =
\left(
\begin{array}{cc}
{\bf 1}_2&0\\
0&-{\bf 1}_2
\end{array}\right)
\end{equation}
which is denoted by $\gamma^{0}$ in Kaluza-Klein theory explicitly involving 
the time \cite{CV2}.

The {\it standard} Dirac operator of the theory without explicit mass term 
is defined as ${D}_{s}=\gamma^{\hat\alpha}\hat\nabla_{\hat\alpha}$  
\cite{CV2,CV3}
where the spin covariant derivatives with local indices, 
$\hat\nabla_{\hat\alpha}$, depend on the momentum operators,  
$P_{i}=-i(\partial_{i}-A_{i}\partial_{4})$ and $P_{4}=-i\partial_{4}$,
and spin connection \cite{CV2},  
such that the  Hamiltonian operator \cite{CV2,CV4},
\begin{equation}\label{HH}
H =\gamma^5 D_{s}
=\left(
\begin{array}{cc}
0&\alpha^{*}\\
\alpha&0
\end{array}\right)
\end{equation}
can be expressed in terms of Pauli operators,  
\begin{eqnarray}
&\alpha&
=\sqrt{V}\left( \vec{\sigma}\cdot\vec{P}-\frac{iP_{4}}{V}\right)\,,\\ 
&\alpha^{*}& 
=V\left(\vec{\sigma}\cdot\vec{P}+\frac{iP_{4}}{V}\right)\frac{1}{\sqrt{V}}\,, 
\end{eqnarray}
involving the Pauli matrices, $\sigma_i$. These  operators give the (scalar) 
Klein-Gordon operator of the 
Taub-NUT space \cite{CV2,CV4},  $\Delta= -\nabla_{\mu}g^{\mu\nu}\nabla_{\nu}=
\alpha^{*}\alpha$. We specify that here the star superscript is a mere 
notation that does not represent the Hermitian conjugation because we are 
using a non-unitary representation of the algebra of Dirac operators. Of 
course, this is {\it equivalent} to the unitary representation where all 
of these operators are self-adjoint \cite{CV2}. 
 
The first three Killing-Yano tensors of the Taub-NUT space \cite{GRFH},  
\begin{equation}\label{fi}
f^i 
= f^i_{\,{\hat \alpha}{\hat \beta}} {\hat e}^{\hat \alpha} \wedge 
{\hat e}^{\hat \beta}
= 2 {\hat e}^4\wedge  {\hat e}^i +\varepsilon_{ijk} {\hat e}^j\wedge 
{\hat e}^k
\end{equation}
are rather special since they are covariantly constant. The $f^i$ 
define three anticommuting complex structures of the Taub-NUT manifold, 
their components realizing the quaternion algebra
\begin{equation}\label{ff}
f^i  f^j + f^j f^i = - 2 \delta_{ij}~~~, ~~~f^i  f^j - f^j f^i = - 2
\varepsilon_{ijk} f^k.
\end{equation}

The existence of these Killing-Yano tensors is linked to the 
hyper-K\"ahler geometry of the manifold and shows directly the relation 
between the geometry and the $N = 4$ supersymmetric extension of the 
theory \cite{GRH,vH1}. Moreover we can give a {\it physical} 
interpretation of these Killing-Yano tensors defining the {\it spin-like} 
operators,
\begin{equation}\label{sl}
\Sigma_{i}=-\frac{i}{4}f^{i}_{\hat\alpha\hat\beta}
\gamma^{\hat\alpha}\gamma^{\hat\beta}=\left(
\begin{array}{cc}
\sigma_i&0\\
0&0
\end{array}\right)\,,
\end{equation}
that have similar properties to those of the Pauli matrices. 
In the pseudo-classical description of a Dirac particle \cite{GRH,vH1}, 
the covariantly constant Killing-Yano tensors correspond to components 
of the spin which are separately conserved.

Here, since the Pauli matrices commute with the Klein-Gordon operator, 
the spin-like operators (\ref{sl}) commute with $H^2$. Remarkable the 
existence of the Killing-Yano tensors allows one to construct {\it 
Dirac-type} operators \cite{CL}
\begin{equation}\label{Qi}
Q_{i}=-if^{i}_{\,\hat\alpha\hat\beta}\gamma^{\hat\alpha}\hat\nabla
^{\hat\beta}=\{H,\,\Sigma_i\}=\left(
\begin{array}{cc}
0&\sigma_i\alpha^{*}\\
\alpha\sigma_i&0
\end{array}\right)
\end{equation}
which anticommute with $D_s$ and $\gamma^5$ and commute with $H$ 
\cite{CV3}. Another Dirac operator can be defined using the fourth 
Killing-Yano tensor but this will be discussed separately in Sec. 5.

\section{Equivalent representations}

In \cite{CV2} we have shown that in the massless case the operators 
$Q_i$ ($i=1,2,3$) and 
the new supercharge $Q_0=iD_s=i\gamma^5 H$ form the basis of
a $N=4$ superalgebra obeying the anticommutation relations   
\begin{equation}\label{QQH}
\{Q_{A},\,Q_{B}\}=2\delta_{AB}H^2\,, \quad A,B,...=0,1,2,3
\end{equation}  
linked to the hyper-K\" ahler geometric structure of the Taub-NUT space. 
In addition, we associate to each Dirac operator $Q_A$ its own 
Hamiltonian operator $\tilde Q_A=-i\gamma^5 Q_A$ obtaining thus another set 
of supercharges,
\begin{equation}\label{ham}
\tilde Q_0=H\,,\quad \tilde Q_i =i[H,\Sigma_i]\,,
\end{equation}  
which obey the same anticommutation relations as (\ref{QQH}). Thus we find 
that there are  two similar  superalgebras of operators with precise 
physical meaning. Obviously, since all of these operators must be 
self-adjoint we have to work only with {\it unitary} representations 
of these superalgebras, up to an equivalence.  
 
The concrete form of these supercharges depends on the representation of 
the Dirac matrices which can be changed at any time with the help of a 
non singular operator $T$ such that all of the $4\times 4$ matrix 
operators of the Dirac theory transform as $X\to X'=TXT^{-1}$. In this 
way one obtains an {\it equivalent} representation which preserves the 
commutation and the anticommutation relations. In \cite{CV2} we have 
used such transformations for pointing out that the convenient 
representations where we work are equivalent to an unitary one. We note 
that some properties of the transformations changing representations in 
theories with two Dirac operators and their possible new applications are 
discussed in \cite{RUSI}. 

In order to produce explicit operators $T$ which connect different 
Dirac operators in what follows we shall consider constant non-singular 
matrices $T$ commuting with 
$H^2={\rm diag}(\Delta, \alpha\,\alpha^{*})$ giving transformations  
$Q_{A}\to Q'_{A}=TQ_{A}T^{-1}$ that lead to  equivalent representations of 
our superalgebra,  $\{Q'_{A},\,Q'_{B}\}=2\delta_{AB}H^2$, with the 
{\it same} $H^2$. Since $\Delta$ is a scalar differential operator while 
$\alpha\,\alpha^{*}$ has complicated spin terms, it is suitable to choose  
matrices of the form 
$T={\rm diag}(\hat T, {\bf 1}_2)$ where $\hat T$ can be any non singular 
$2\times 2$ constant matrix. 
In general, these transformations lead to new supercharges $Q'_{A}$ which 
are linear combinations of the original ones with mixing coefficients that 
can be complex numbers. The basic principles of quantum mechanics require 
the Dirac-type operators to be self-adjoint (up to an equivalence)  as the 
standard Dirac operator \cite{CL}. Therefore, if one starts with a suitable 
representation then it is recommendable to use only  {\it unitary} 
transformations of the form
\begin{equation}
U(\beta,\vec{\xi})=\left(
\begin{array}{cc}
\hat U(\beta,\vec{\xi})&0\\
0&{\bf 1}_{2}
\end{array}\right)\,,
\end{equation}
where $\hat U(\beta, \vec{\xi})=e^{-i\beta}\hat U(\vec{\xi})
\in U(2)=U(1)\otimes SU(2)$ with $\hat U(\vec{\xi}) \in SU(2)$. This is 
because among these transformations one could find those linking 
equivalent Dirac operators.  

It is interesting to observe that the $SU(2)$ transformations are 
generated just by the above defined spin-like operators as 
\begin{equation}\label{Uxi}
U(\vec{\xi})=U(0,\vec{\xi})=e^{-i\vec{\xi}\cdot \vec{\Sigma}/2}=\left(
\begin{array}{cc}
\hat U(\vec{\xi})&0\\
0&{\bf 1}_{2}
\end{array}\right)\,,
\end{equation}
If we take now $\vec{\xi}=2\varphi\, \vec{n}$ with $|\vec{n}|=1$ and 
$\varphi \in [0, \pi]$, we find that 
\begin{equation}
\hat U(\vec{\xi})=e^{-i\vec{\xi}\cdot\vec{\sigma}/2}=
{\bf 1}_2 \cos\varphi -i\vec{n}\cdot\vec{\sigma}\sin\varphi
\end{equation}
and after a little calculation we can write the concrete action of 
(\ref{Uxi})  as
\begin{eqnarray}
Q'_{0}&=&U(\vec{\xi})Q_{0}U^{+}(\vec{\xi})=Q_{0}\cos\varphi + 
n_i Q_{i}\sin\varphi \label{Q1}\\   
Q'_{i}&=&U(\vec{\xi})Q_{i}U^{+}(\vec{\xi})=Q_{i}\cos\varphi - 
\left(n_i Q_{0}+\varepsilon_{ijk}n_{j}Q_{k}\right)\sin\varphi \label{Q2}\,.   
\end{eqnarray}
Hereby we see that the supercharges are mixed among themselves in linear 
combinations involving only {\it real} coefficients. In addition, we 
observe that these transformations correspond to an irreducible 
representation since the supercharges transform like the real components 
of a Pauli spinor. In other words, the usual $SU(2)$ transformations 
$\psi_{Q}\to \psi_{Q}'=\hat U^{+}(\vec{\xi})\psi_{Q}$ 
of the spinor-operator  
\begin{equation}
\psi_{Q}=\left(
\begin{array}{c}
Q_0 -iQ_{3}\\
Q_{2}-iQ_{1}
\end{array}\right)
\end{equation}
give just the  transformations (\ref{Q1}) and (\ref{Q2}).  
 
\section{Discrete transformations}

Let us focus now only on the  transformations which transform the 
supercharges $Q_A$ among themselves without to affect their form. 
{}From (\ref{Q1}) and (\ref{Q2}) we see that there exists particular 
transformations, 
\begin{equation} 
Q_k=U_k Q_0 U^{+}_k\,, \quad k=1,2,3\,, 
\end{equation}
where the matrix $U_k={\rm diag} (-i\sigma_{k},{\bf 1}_2)$ is given by 
$-i\sigma_k \in SU(2)$.   
In addition, we consider the {\it parity} operator $P=P^{-1}=-\gamma^5$ 
which changes the sign of supercharges, 
\begin{equation}\label{par}
PQ_A P=-Q_A\,, \quad A=0,1,2,3\,. 
\end{equation}
Then it is not hard to verify that the 
identity $I={\bf 1}_4$,  $P$ and the  sets of matrices $U_{k}$ and 
$PU_k$ ($k=1,2,3$) form a discrete group of order eight 
the multiplication table of which is determined by the following rules 
\begin{eqnarray}\label{mult}
&&P^2=I\,, \quad PU_k=U_k P\,,\nonumber\\
&&{U_1}^2={U_2}^2={U_3}^2=P\,,\label{A}\\
&&U_1 U_2 =U_3\,,\quad U_2 U_1 =PU_3\,,...\,{\rm etc.}\nonumber
\end{eqnarray}
We denote this group by ${\cal G}_Q$ since it is a realization of the 
quaternion group ${\bf Q}$ which is isomorphic with the dicyclic group 
$\left<2,2,2\right>$ \cite{BR,CM} (see the Appendix). In the 
representation (\ref{(gammai)}) of the $\gamma$-matrices, its operators 
are defined by proper unitary matrices (which satisfy $G^{-1}=G^{+}$ and 
${\rm det} G=1\,, \forall\, G\in {\cal G}_Q)$ constructed using the 
elements $\pm{\bf 1}_2, \pm i\sigma_1, \pm i\sigma_2, \pm i\sigma_3$ of 
the natural realization of ${\bf Q}$ as a discrete subgroup of $SU(2)$. 

The group  ${\cal G}_Q$ is interesting because it brings together the 
parity that produces the transformations (\ref{par}) and the operators 
$U_k$ giving sequences of the form 
\begin{equation}
Q_1=U_3^{+}Q_2 U_3=U_2 Q_3 U_2^{+}=U_1 Q_0 U_1^{+}, ... {\rm etc.}
\end{equation}
which lead to the conclusion that the Dirac operators and their parity 
transformed, $\pm Q_A$ $(A=0,1,2,3$), are equivalent among themselves. All 
these operators constitute the  orbit 
$\Omega_Q=\{Q\,|\,Q=GQ_{0}G^{+}\,,\forall G\in {\cal G}_Q\}$
of the group ${\cal G}_Q$ in the algebra of the $4\times 4$ matrix 
operators. A similar orbit, $\tilde\Omega_Q$, can be constructed for 
the associated  Hamiltonian operators, $\pm\tilde Q_A$ defined by 
(\ref{ham}), if we start with $\tilde Q_0$ instead of $Q_0$. It is 
remarkable that  each of these two orbits includes {\em only} 
operators representing (up to sign) supercharges obeying superalgebras 
of the form (\ref{QQH}).

In the Kaluza-Klein theory with the time trivially added \cite{CV2}, the 
time dependent term of the whole massless Dirac operator  commutes with 
all the operators of ${\cal G}_Q$ such that it remains unchanged when 
one replaces the space parts using the discrete transformations of this 
group. In these conditions  all the Dirac operators of $\Omega_Q$ lead 
to equivalent Dirac equations from the physical point of view. 
These  can be written in Hamiltonian form as 
$i\partial_t \psi_A^{(\pm)}=\pm\tilde Q_A \psi_A^{(\pm)}$ ($A=0,1,2,3$) 
and produce the same energy spectrum which coincides to that of the 
Klein-Gordon equation as it results  from the superalgebra 
(\ref{QQH}) \cite{CH,CV2}. 
         
The existence of this discrete symmetry among the four supercharges of 
the superalgebra of the Dirac and Dirac-type operators (or the 
corresponding Hamiltonian operators) must be understood as a consequence 
of the fact that the Taub-NUT space has a hyper-K\" ahler structure 
modeled on a quaternion inner-product space \cite{H}. In other words, 
the Dirac theory in this space picks up the basic quaternion character of 
the tangent space showing it off  as the  discrete symmetry due to the 
group ${\cal G}_Q\sim {\bf Q}$,  naturally related to the specific 
supersymmetries of this geometry.    

\section{Hidden symmetries and the fifth Dirac \\ operator}

In the Taub-NUT space, in addition to the above discussed 
covariantly constant Killing-Yano tensors, there exists a fourth 
Killing-Yano tensor,  
\begin{equation}\label{fy}
f^Y 
= - \frac{x^i}{r}f^i +\frac{2 x^i}{\mu V}\varepsilon_{ijk} {\hat e}^j\wedge 
{\hat e}^k\,,
\end{equation}
which is not covariantly constant. The presence of $f^Y$ is due to the 
existence of the hidden symmetries of the Taub-NUT geometry which are 
encapsulated in three non-trivial St\"ackel-Killing tensors. These 
are interpreted as the components of the so-called Runge-Lenz vector of 
the Taub-NUT problem and are expressed as symmetrized products of the 
Killing-Yano tensors $f^Y$ and $f^i,\, (i=1,2,3)$. 

As in the case of the Dirac operators (\ref{Qi}), one can 
use $f^Y$ for defining the fifth Dirac operator 
\begin{equation}
Q^{Y}_0=-i \gamma^{\hat\alpha}\left(
f^{Y}_{\,\hat\alpha\hat\beta}\hat\nabla^{\hat\beta}
-\frac{1}{6}\gamma^{\hat\beta}\gamma^{\hat\delta}
f^{Y}_{\,\hat\alpha\hat\beta;\hat\delta}
\right)\,,
\end{equation}
called here the {\it non-standard} or {\it hidden} Dirac operator to 
emphasize the connection with the hidden symmetry of the Taub-NUT 
problem. It is denoted by $Q_0^Y$ instead of $Q^Y$ as in \cite{CV3} 
to point out its relation to the standard Dirac operator since it can 
be put in the form
\begin{equation}
Q^Y_0=i\frac{r}{\mu}\left[  Q_0\,,\left(
\begin{array}{cc}
\sigma_r&0\\
0&\sigma_r V^{-1}
\end{array} \right) \right]\,,
\end{equation}
where $\sigma_r=\vec{x}\cdot \vec{\sigma}/r$. We showed that  
$Q^Y_0$ commutes with $\tilde Q_{0}=H$ and anticommutes with $Q_0$ and 
$\gamma^5$ \cite{CV3}. This operator is important because it allowed us 
to derive  the explicit form of the Runge-Lenz operator, $\vec{K}$, of 
the Dirac field in Taub-NUT background establishing its properties 
\cite{CV3}. We recall that the components of the conserved total angular 
momentum, $\vec{J}$, and the operators $R_i=F^{-1}K_i$ with 
$F^2={P_4}^2-H^2$ are just the generators of the dynamical algebra of 
the Dirac theory in Taub-NUT background \cite{CV4}.

Starting with $Q^Y_0$ we can construct a new orbit,  
$\Omega^Y$, of ${\cal G}_Q$ defining 
\begin{equation}\label{QYk}
Q^{Y}_k=U_k Q_0^YU_k^{+}=
i\frac{r}{\mu}\left[ Q_k\,,\left(
\begin{array}{cc}
\sigma_k \sigma_r\sigma_k&0\\
0&\sigma_r V^{-1}
\end{array} \right) \right]
\end{equation}
(for $k=1,2,3$) and observing that
\begin{equation}
P Q^Y_A P=-Q^Y_A\,, \quad  A=0,1,2,3\,.
\end{equation}
{}From the explicit form (\ref{QYk}) we deduce that, in contrast with the 
operators of the orbits $\Omega_Q$ and $\tilde\Omega_Q$, those of the orbit 
$\Omega^Y$ have more involved algebraic properties. We can convince that 
calculating, for example, the identity
\begin{equation}\label{HQ2}
H^2(Q^Y_0)^2=H^4+\frac{4}{\mu^2}H^2\left(\vec{J}^2+\frac{1}{4}\right)+
4F^2{P_4}^2\,,
\end{equation}
and it is worth comparing it with equation (\ref{QQH}). The Dirac-type 
operators $Q_A$ are characterized by the fact that their quantal 
anticommutator close on the square of the Hamiltonian of the theory. No 
such expectation applies to the non-standard, hidden Dirac operators 
$Q^Y_A$ which close on a combination of different conserved operators.
Also from equation(\ref{HQ2}) it results that  $(Q_A^Y)^2 \not =(Q_B^Y)^2$ 
if $A\not= B$ (because $\vec{J}^2$ does not commute with $U_k$). Moreover, 
one can show that the commutators $[Q^Y_A,\,Q^Y_B]$ have complicated forms 
which can not be expressed in terms of operators $Q_A^Y$. Therefore, 
neither the commutator nor the anticommutator of the pairs of operators of 
this orbit do not lead to significant algebraic results as the 
anticommutation relations (\ref{QQH}) of the operators 
$Q_A , \, (A = 0,1,2,3)$.

Thus we conclude that the operators of the orbit $\Omega^Y$ do not form a 
closed algebraic structure. The unique virtue of the equivalent operators 
$\pm Q^Y_A$ is that they commute with the corresponding Hamiltonian 
operators $\tilde Q_A\,,(A=0,1,2,3)$. In this way we see that the discrete 
symmetry given by ${\cal G}_Q$ is {\em decoupled} from the hidden 
symmetries which have a different geometric origin. Its existence requires 
the Weyl tensor to be of Petrov type D. For this reason it is pointless to 
use the whole orbit $\Omega^Y$, the operator $Q^Y_0$ being enough for 
deriving the components of the Runge-Lenz operator.

\section{Discussion}

In this article we  pointed out the existence the discrete symmetry group 
${\cal G}_Q \sim {\bf Q}$ of the Dirac theory in Taub-NUT space 
which plays here the same role as the simpler discrete group 
${\bf Z}_2\subset {\bf Q}$ of the usual theory of the Dirac field in 
Minkowski background (formed only by the identity and parity operators). 
The operator $P\in {\cal G}_Q$ is interpreted as the parity operator in 
the massless case of the Kaluza-Klein theories with the time trivially 
added because it is a proper transformation which  changes the sign of 
the space part of the Dirac operator (or of the Hamiltonian one). 
In other theoretical conjectures the interpretation of $P$ can be 
different. One of the first examples given of a gravitational instanton 
was the self-dual Taub-NUT solution \cite{SH}. The gravitational instantons 
are complete non-singular Einstein metrics, usually taken to have $(++++)$ 
signature. For this reason in these theories the  operator 
$P$ is interpreted as the {\it TP} reversal, changing the sign of 
all the coordinates $x^\mu$. 

As it is expected the parity operator $P$ is involved in the relation 
between the indices of the Dirac-type operators. Taking into account 
relations (\ref{QQH}) and (\ref{par}) we observe that any pair
of operators  $(Q_A , PQ_B)$ with $A\not=B$ can always be diagonalized 
simultaneously. Hereby it results that the kernels of all four Dirac-type 
operators coincide. In even-dimensional spaces the index of a Dirac operator 
can be defined as the difference in the number of linearly independent zero 
modes with eigenvalues $+1$ and $-1$ under $\gamma^5$. It is quite 
simple to get the remarkable result that the index of all Dirac 
operators is the same \cite{RUSI,HWP}. An immediate consequence is that 
the operators $Q_A$ have the same zero-modes. However  
the zero-modes of the fifth non-standard Dirac operator $Q^Y$ coincide 
with those of the other Dirac-type operators only in some peculiar
cases \cite{vH1} even though the index of the operator 
$Q^Y$ is equal to the index of the Dirac-type operators $Q^A$ 
\cite{HWP}.

In conclusion we can say that the Taub-NUT space has a special geometry 
where the covariantly constant Killing-Yano tensors exist by 
virtue of the metric being self-dual and the Dirac-type operators 
generated by them are equivalent with the standard one. All of these 
operators which form the orbit $\Omega_Q$ of ${\cal G}_Q$   
accomplish the anticommutation relation (\ref{QQH}). The fourth 
Killing-Yano tensor $f^Y$ which is not covariantly constant exists by 
virtue of the metric being of type $D$. The corresponding non-standard 
or hidden Dirac operator does not close on $H$ as it can be seen from 
equation (\ref{HQ2}) and is not equivalent to the Dirac-type operators. 
As it was mentioned, it is 
associated with the hidden symmetries of the space allowing the 
construction of the conserved vector-operator analogous to the Runge-Lenz 
vector of the Kepler problem.  
Here we have shown how the discrete symmetry given by ${\cal G}_Q$ is 
naturally related only to the supersymmetries, being decoupled from the   
hidden symmetries which have another geometric source.  

\subsection*{Acknowledgments}

One of us (M.V.) would like to thank Institute for Theoretical Physics, 
Bern, Switzerland for the hospitality extended to him while part of 
this work was performed. Useful discussions with J. Gasser, P. Hajicek
and P. Minkowski are also acknowledged.

\setcounter{equation}{0} \renewcommand{\theequation}
{A.\arabic{equation}}

\section*{Appendix: The quaternion group}

The dicyclic group $\left<2,2,m\right>$ of order $4m$ is defined as the 
discrete group generated by two elements, $x$ and $y$, obeying
\begin{equation}
x^4=e\,,\quad x^2=y^m\,, \quad yx=xy^{-1}
\end{equation}
where $e$ is the unit element \cite{CM}. 

For $m=2$ we denote $u_1=x$, $u_2=y$ and $u_3=xy$ finding that the new 
element $p={u_1}^2 ={u_2}^2 ={u_3}^2$ satisfies $p^2=e$. If, in addition, 
we consider the elements  $pu_1$, $pu_2$ and $pu_3$, we recover similar 
multiplication rules as (\ref{mult}). On the other hand, taking 
$e=1$, $p=-1$ and  $u_1, u_2, u_3$ the quaternion complex constants one 
generates the quaternion group ${\bf Q}\sim \left<2,2,2\right>$. We recall 
that the pair $(e\,,p)\sim (1,-1)$ forms the cyclic group 
${\bf Z}_2\subset {\bf Q}$.

\end{document}